\documentclass[journal]{IEEEtran}

\usepackage{amsmath,amssymb,amsfonts,mathrsfs,mathtools,dsfont,mathrsfs, amsthm, bm, bbm,pdfsync,xfrac}
\DeclareMathAlphabet{\mathpzc}{OT1}{pzc}{m}{it}
\usepackage{csvsimple,booktabs}
\usepackage{float,enumitem,subfig}
\usepackage{graphicx,epsfig,xcolor}
\usepackage{multirow}

\usepackage[hyphens]{url}
\usepackage[breaklinks, colorlinks, linkcolor=purple, anchorcolor=purple, citecolor=purple, urlcolor=purple]{hyperref}
\usepackage{cite}

\usepackage{algorithm,algpseudocode}

\newcommand{\beq}{\begin{equation}}
\newcommand{\eeq}{\end{equation}}
\newcommand{\beqa}{\begin{eqnarray}}
\newcommand{\eeqa}{\end{eqnarray}}
\newcommand{\beqan}{\begin{eqnarray*}}
\newcommand{\eeqan}{\end{eqnarray*}}

\newcommand{\vnorm}[1]{\left\|#1\right\|}

\newcommand{\E}{\mathds{E} }

\newcommand{\prob}{\mathbb{P}}

\newcommand\T{{\mathpalette\raiseT\intercal}}
\newcommand\raiseT[2]{\raisebox{0.25ex}{$#1#2$}
}

\newcommand{\conv}{\textrm{C}}

\newcommand{\Aset}{\mathbb{A}}

\newcommand{\Iset}{\mathbb{I}}

\newcommand{\Rset}{\mathbb{R}}

\newcommand{\Uset}{\mathbb{U}}

\newcommand{\Acal}{{\cal A}}
\newcommand{\Bcal}{{\cal B}}

\newcommand{\Dcal}{{\cal D}}
\newcommand{\Ecal}{{\cal E}}

\newcommand{\Gcal}{{\cal G}}

\newcommand{\Kcal}{{\cal K}}
\newcommand{\Lcal}{{\cal L}}

\newcommand{\Scal}{{\cal S}}

\newcommand{\Vcal}{{\cal V}}

\newcommand{\Gfk}{{\mathfrak{G}}}

\newcommand{\Ascr}{\mathscr{A}}

\newcommand{\Hscr}{\mathscr{H}}

\newcommand{\Oscr}{\mathscr{O}}

\newcommand{\bone}{\mathds{1}}

\renewcommand{\v}[1]{{\bm{#1}}}

\newcommand{\ol}[1]{\ensuremath{\overline{{#1}}}}
\newcommand{\ul}[1]{\ensuremath{\underline{{#1}}}}

\newcounter{l1}
\newcounter{l2}
\newcounter{l3}
\newcommand{\bdotlist}{\begin{list}{$\bullet$}{}}
\newcommand{\bboxlist}{\begin{list}{$\Box$}{}}
\newcommand{\bbboxlist}{\begin{list}{\raisebox{.005in}{{\tiny
$\blacksquare$ \ \ }}}{}}
\newcommand{\bdashlist}{\begin{list}{$-$}{} }
\newcommand{\blist}{\begin{list}{}{} }
\newcommand{\barablist}{\begin{list}{\arabic{l1}}{\usecounter{l1}}}
\newcommand{\balphlist}{\begin{list}{(\alph{l2})}{\usecounter{l2}}}
\newcommand{\bAlphlist}{\begin{list}{\Alph{l2}.}{\usecounter{l2}}}
\newcommand{\bdiamlist}{\begin{list}{$\diamond$}{}}
\newcommand{\bromalist}{\begin{list}{(\roman{l3})}{\usecounter{l3}}}

\newcommand{\CVaR}{{\rm CVaR}}
\newcommand{\VaR}{{\rm VaR}}
\newcommand{\PHCA}{{\rm PHCA}}

\newcommand{\hadamard}{\odot}

\newcommand{\vpd}{\v{p}_{\textrm{D}}}
\newcommand{\vqd}{\v{q}_{\textrm{D}}}
\newcommand{\vpg}{\v{p}_{\textrm{G}}}
\newcommand{\vqg}{\v{q}_{\textrm{G}}}
\newcommand{\vpinj}{\v{p}_{\textrm{inj}}}
\newcommand{\vqinj}{\v{q}_{\textrm{inj}}}
\newcommand{\etaG}{\v{\eta}_{\textrm{G}}}
\newcommand{\sample}{\textrm{sample}}

\newcommand{\KSO}{\Kcal_{\textrm{SO}}}
\renewcommand\top{{\mathpalette\raiseT\intercal}}
\renewcommand\raiseT[2]{\raisebox{0.25ex}{$#1#2$}
}

\begin{document}

\title{Conditional Value at Risk-Sensitive Solar Hosting Capacity Analysis in Distribution Networks}

\author{Avinash N. Madavan \qquad Nathan Dahlin \qquad Subhonmesh Bose \qquad Lang Tong
\thanks{A. N. Madavan, N. Dahlin, and S. Bose are with the Department of Electrical and Computer Engineering at the University of Illinois at Urbana-Champaign, Urbana IL, USA 61801. L. Tong is with the School of Electrical and Computer Engineering at Cornell University, Ithaca NY, USA 14680. This work was partially supported by the NSF CAREER grant 2048065. Emails: \{madavan2, dahlin, boses\}@illinois.edu, lt35@cornell.edu}}

\maketitle

\begin{abstract}
Solar hosting capacity analysis (HCA) assesses the ability of a distribution network to host distributed solar generation without seriously violating distribution network constraints. In this paper, we consider risk-sensitive HCA that limits the risk of network constraint violations with a collection of scenarios of solar irradiance and nodal power demands, where risk is modeled via the conditional value at risk (CVaR) measure. First, we consider the question of maximizing aggregate installed solar capacities, subject to risk constraints and solve it as a second-order cone program (SOCP) with a standard conic relaxation of the feasible set with power flow equations. Second, we design an incremental algorithm to decide whether a configuration of solar installations has acceptable risk of constraint violations, modeled via CVaR. The algorithm circumvents explicit risk computation by incrementally constructing inner and outer polyhedral approximations of the set of acceptable solar installation configurations from prior such tests conducted. Our numerical examples study the impact of risk parameters, the number of scenarios and the scalability of our framework.
\end{abstract}


\section{Introduction}

According to a recent study in \cite{joshi2021high}, usable rooftop space in the United States can produce enough solar energy to meet the nation's current electricity demands. In 2022, new utility-scale solar is projected to account for almost half of all planned capacity expansions, per the Energy Information Administration. 
Although small-scale distributed rooftop solar (with capacities less than 1MW) only produces about 1\% of total electricity generation today, this sector is rapidly growing, driven in part by ambitious targets set forth by renewable portfolio standards (e.g., in \cite{hawaii100}). California has exceeded 10GW of installed capacity of distributed solar in 2022. Distributed solar accounts for more than $60$\% of solar projects in New York, Hawaii, Maryland, New Jersey and Massachusetts. These numbers are expected to grow, aided by legislation such as Title 24 in California and bill S2165 in Massachusetts that require new building constructions to install rooftop solar. Given this rapid growth, we ask: \emph{is the grid ready to host distributed solar at scale}? 

Solar production curtailments, exceeding 15\% of total potential output, have often been necessary in California to avoid violation of system constraints, including voltage and harmonic ranges, and protection system failures, etc. See \cite{EIACAISOcurtail,azibek2020hosting} for details. Today, these interventions  impact utility-scale solar installations. However, similar questions arise for distributed solar in low/medium-voltage distribution networks. Specifically, how much installed solar capacity can the distribution network host, without jeopardizing safe operational limits for the network? Such quantification studies are categorized as \emph{hosting capacity analyses} (HCA). This paper aims to provide efficient mathematical frameworks for the same.

Traditional approaches to HCA have relied on historic peak load levels or point estimates of solar generation and network load profiles, e.g., see \cite{SGIP,baldenko2016determination}. Distributed solar is highly intermittent (especially during the spring season), and network load characteristics are expected to change significantly with the increased adoption of new technologies such as home batteries and electric vehicles. Consequently, deterministic HCA, conducted with a few representative scenarios, will fail to capture the variety of operating conditions the network might face. See \cite{SGIP,capitanescu2014assessing} for a discussion. Instead, one must account for the \emph{statistics} of operating conditions, both observed in historical data and those obtained with reasonable projections of load characteristics into the future.  Several recent works have indeed proposed to consider said statistics in HCA through stochastic programming; see \cite{geng2020probabilistic} for a survey.

\begin{figure}[!ht]
  \centering
  \vspace{-0.1in}
  \includegraphics[width=0.94\linewidth]{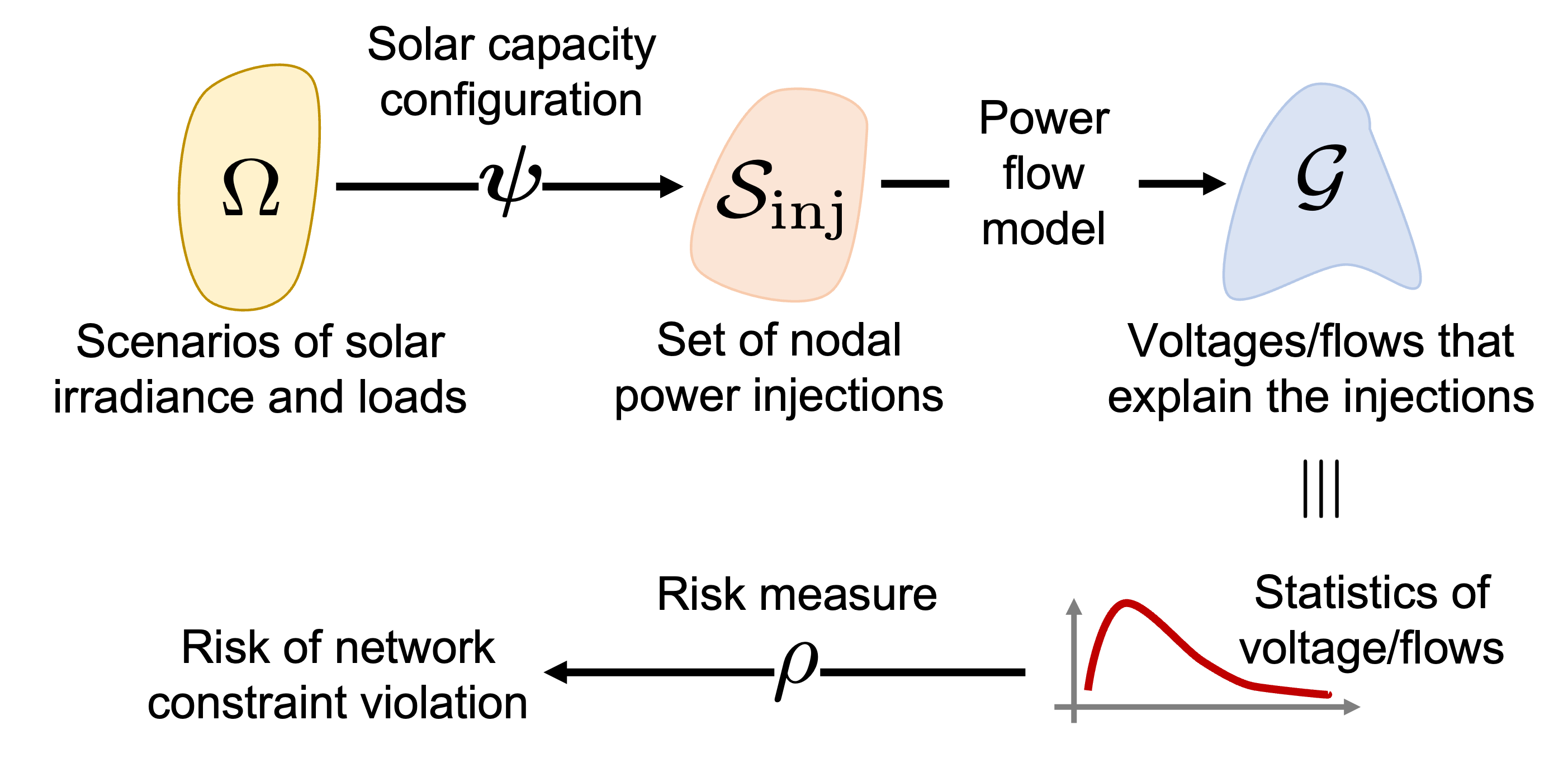}
  \caption{The work-flow of risk-sensitive solar HCA.}
  \label{fig:HCA}
\end{figure}

The central challenge of stochastic optimization-based HCA is \emph{scalability}. Such techniques must consider large numbers of possible grid scenarios, drawn from sampled or synthesized data. Using parametric families of distributions to describe such data is often difficult, especially if synthetic data is obtained from generative models. HCA methods that use scenario-based optimization typically entail the solution of a large number of power flow problems to compute network-relevant variables such as voltages and power flows in each scenario. Then, the  statistics of possible network constraint violations with these voltages/power flows are processed, possibly through a risk measure, e.g., the probability of constraint violations in \cite{geng2020probabilistic}. See Figure \ref{fig:HCA} for an illustration of the work-flow of stochastic HCA methods.
The key innovation of our work lies in the use of the \emph{conditional value at risk} (\CVaR{}) measure to process the statistics of constraint violation, as we explain in Section \ref{sec:HCA_PCA}. While the use of this risk measure is popular in finance (e.g., see \cite{kisiala2015conditional,lim2011conditional}), its favorable mathematical properties are popularizing its use in engineering domains, including power systems. 
As will become evident, these properties allow efficient algorithm design to tackle the scalability challenge of HCA.


The specific contributions of this paper are as follows. First in Section \ref{sec:formulation}, we propose a \CVaR{}-sensitive hosting capacity maximization problem that seeks to identify the maximum aggregate capacity of solar installations in a distribution network, and describe a scenario-based approach to solve it in Section \ref{sec:scenario} via \emph{convex optimization}. This convexity follows from our use of \CVaR{}, as well as a standard second-order cone programming-based relaxation of the DistFlow power flow model. The vast majority of risk measures introduce nonconvexity. Moreover, such measures require explicit evaluation of the risk of constraint violations across scenarios in order to assess the risk of a particular installation configuration. As a result, one must consider solar installations one at a time, assess the risk level for each, and update the configuration suitably to optimize it. In contrast,  \CVaR{} preserves the convexity of our power flow model and allows for formulation of an optimization problem that directly optimizes over realizable installations according to risk level, defined through scenarios, and quality as captured by the chosen objective function. Convexity allows us to leverage mature off-the-shelf convex optimization tools and avoids the pitfalls of constraint linearization, mixed integer programming, or the design of complex, parameter-sensitive algorithms to explore the space of capacity configurations that require careful parameter-tuning, e.g., in \cite{ding2016distributed,abad2018probabilistic,santos2016new}. 
Our numerical experiments demonstrate the impact of risk parameters and the number of scenarios on the solution quality and runtime.

Second, we propose a method in Section \ref{sec:feasibility} to quickly determine whether a candidate solar installation configuration has acceptable risk of network constraint violations, as evaluated via \CVaR{}. 
Our proposed  method is incremental, i.e., we utilize the results of prior tests to expedite 
determination as to whether a new candidate capacity configuration is acceptable. The properties of the \CVaR{} measure dictate that the resulting set of acceptable configurations is convex. Thus, we construct and refine convex inner and outer approximations to this set as more configurations are tested. These approximations often permit us to  quickly certify whether a new candidate configuration is acceptable, without having to explicitly test for it using all scenarios. Our numerical examples illustrate that acceptability certification time decreases rapidly, as the knowledge builds over time from prior tests.
We end the paper in Section \ref{sec:conclusion} with concluding remarks.

\section{CVaR-Sensitive HCA Problems}\label{sec:HCA_PCA}

Solar HCA assesses whether a distribution network can support distributed solar generation from installed capacity configurations without seriously violating distribution network constraints.
In this section, we present abstract formulations of the \CVaR{}-sensitive solar HCA problems that we aim to study in the rest of the paper, utilizing Figure \ref{fig:HCA}. 

Consider a set of solar irradiance and load scenarios $\Omega$, i.e., $\Omega$ captures all uncertainty that the network faces from solar productions and consumer power demands. For a scenario $\omega\in\Omega$, its solar irradiance $\v{\alpha}(\omega)$ translates into solar power generation across the network via a configuration of installed solar capacities $\v{\psi}$. For example, each entry of $\v{\psi}$ might specify the surface area of rooftop photovoltaic panels at the network buses. 
Accompanying the conditions affecting solar generation, scenarios also include power demands at each bus. Thus, scenarios in $\Omega$ and solar capacity configuration $\v{\psi}$ yield a set of corresponding net nodal power injections $\Scal_{\textrm{inj}}$, identified as solar generation less power demands at each bus. These injection scenarios then serve as inputs to a power flow model, which identifies a set $\Gcal$ of voltages and power flows throughout the network that are consistent with the real/reactive power injections in $\Scal_{\textrm{inj}}$. 

For deterministic hosting capacity analysis, $\Omega$ is a singleton. In such an analysis, an installed capacity configuration $\v{\psi}$ is deemed acceptable, if there exist voltages and power flows over the distribution network, given the power flow model, that satisfy a collection of network constraints, e.g., thermal limits, voltage limits, etc. For the risk-sensitive counterpart, we do not certify the acceptability of $\v{\psi}$, based on a single scenario $\v{\omega}\in\Omega$, but do so \emph{across} $\Omega$. Thus in the bottom portion of Figure \ref{fig:HCA}, we consider the \emph{statistics} of the voltages and power flows arising from the scenarios in $\Omega$ under an installation configuration $\v{\psi}$. We utilize a risk-measure $\rho$ to process the statistics of network constraint violations, and call $\v{\psi}$ feasible, if the ``risk'' of network constraint violation is low enough. In this framework, acceptability is determined in a statistical sense, and we consider two questions:
\begin{itemize}[leftmargin=*, nolistsep]
\item $\Ascr(\v{\psi})$: Decide whether an installation configuration $\v{\psi}$ is acceptable.
\item $\Oscr$: Optimize a suitable metric of installed capacities $\v{\psi}$ over acceptable configurations.
\end{itemize}

\begin{figure}[h!]
  \centering
  \includegraphics[width=0.70\linewidth]{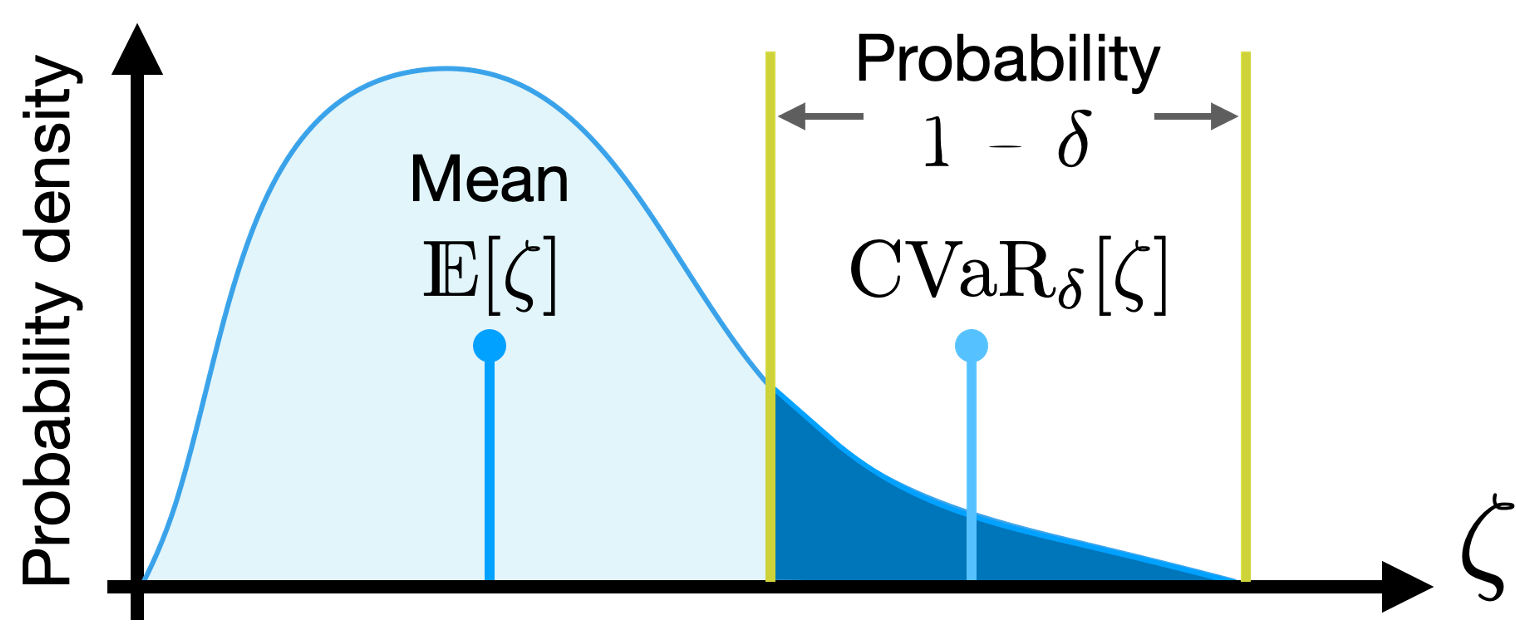}
  \caption{Mean and the conditional value at risk (\CVaR{}) at level $\delta$ for a random variable $\zeta$.}
  \label{fig:CVaR}
\end{figure}
In this work, we use the \emph{conditional value at risk} (\CVaR{}) measure for $\rho$. 
For a scalar random variable $\zeta$, we have
\begin{equation}\label{eq:cvar_variational}
    \CVaR{}_{\delta}[\zeta] := \underset{t\in\Rset}{\text{minimum}}\left\{t + \frac{1}{1-\delta}\E[[\zeta-t]_+]\right\},
\end{equation}
given a parameter $\delta\in[0,1)$. 
Here, $\E[\cdot]$ computes the expectation with respect to the probability distribution over $\zeta$. We use the notation $[z]_+ = \max\{0, z\}$ for any scalar $z$. Driving  $\delta\downarrow 0$, $\text{\CVaR{}}_{\delta}[\zeta]$ approaches the mean value of $\zeta$. For $\delta\uparrow 1$, $\text{\CVaR{}}_{\delta}[\zeta]$ becomes the highest value (essential supremum) that $\zeta$ can take. For a random variable $\zeta$ that admits a smooth cumulative distribution function $F_\zeta$, as in Figure \ref{fig:CVaR}, \CVaR{} gives the expectation over the $(1-\delta)$-tail of the distribution, i.e.,
\begin{equation}\label{eq:cvar_cont}
    \CVaR{}_{\delta}[\zeta] := \mathbb{E}[\zeta\,|\,\zeta\geq F_\zeta^{-1}(\delta)],
\end{equation}
The definition in \eqref{eq:cvar_variational} applies to random variables with arbitrary distributions. See \cite{rockafellar2002conditional} for details. Thus, given an installation configuration $\v{\psi}$, $\Ascr(\v{\psi})$ seeks to decide, if
\begin{align}
    \CVaR_\delta[{g}_i(\v{\psi}; \Omega)] \leq 0, \quad i=1,\ldots,I,
    \label{eq:P1.abstract}
\end{align}
where $g_i$ encodes the network constraints on voltages and power flows that support the power injections, given solar capacities $\v{\psi}$ and a scenario $\omega \in \Omega$. Out of all acceptable configurations, $\Oscr$ identifies those which optimize a particular objective:
\begin{align}
    \text{maximize} \ \text{aggregate capacity } f(\v{\psi}),\
    \text{subject to }  \eqref{eq:P1.abstract}.
\label{eq:P2.abstract}
\end{align}
One can show that, as Figure \ref{fig:CVaR} suggests, 
\begin{align}
    \CVaR{}_{\delta}[g_i(\v{\psi}; \Omega)]\leq 0\implies \prob\{g_i(\v{\psi}; \Omega) > 0 \}\leq 1-\delta.
\end{align}
In other words, the \CVaR{}-sensitive constraint in \eqref{eq:P1.abstract} automatically limits the violation probability of network constraints. Additionally, it also tightly controls the  \emph{extent} of such violations. \CVaR{} has emerged as a popular tool for risk management in finance and engineering, e.g., see \cite{lim2011conditional, kisiala2015conditional}. See \cite{bruno2016risk,madavan2019risk} for its recent use in power system operation. As will become evident, several properties of $\CVaR$ make it particularly suitable to our problem setting. 

In Section \ref{sec:formulation}, we define the distribution network model and present concrete formulations of $\Ascr(\v{\psi})$ and $\Oscr$ as optimization problems. These problems are nonconvex, owing to Kirchhoff's laws. Then, we \emph{convexify} them to $\Ascr^\conv(\v{\psi})$ and $\Oscr^\conv$, by considering a standard second-order cone programming-based relaxation of the feasible set that is defined by the power flow equations. In Section \ref{sec:scenario}, we develop a scenario-based counterpart $\Oscr^\conv_\sample$ of $\Oscr^\conv$ by exploiting the properties of $\CVaR$. In Section \ref{sec:feasibility}, we propose an incremental algorithm to solve the scenario-based variant $\Ascr^\conv_\sample(\v{\psi})$ of $\Ascr^\conv(\v{\psi})$ to decide acceptability of $\v{\psi}$.

\section{Formulating \texorpdfstring{$\Ascr, \Oscr$}{A, O} and their Convex Relaxations \texorpdfstring{$\Ascr^\conv, \Oscr^\conv$}{Ac, Oc}}
\label{sec:formulation}


We begin by encoding the single-phase equivalent of a three-phase radial distribution network as a directed graph $\Gfk(\Vcal, \vec{\Ecal})$. Let $\Vcal = \{1, \ldots, n\}$ describe the distribution buses, where the first bus is the substation. We say $j \to j' \in \vec{\Ecal}$, if bus $j$ is connected to bus $j'$ via a distribution line. Assign the directions of these edges arbitrarily. Given that the radial network $\Gfk$ has $n-1$ edges, the edge-to-node adjacency matrix $\v{B} \in \Rset^{(n-1) \times n}$ is described by
\begin{align}
    \v{B}_{ej} = 
    \begin{cases}
    +1, & \text{if } e = j \to j' \in \vec{\Ecal} \text{ for some } j' \in \Vcal, 
    \\
    -1, & \text{if } e = j' \to j \in \vec{\Ecal} \text{ for some } j' \in \Vcal, 
    \\
    0, & \text{otherwise}.
    \end{cases}
\end{align}

Consider a probability space $(\Omega, \Bcal, \prob)$, where $\Omega$ encodes all possible scenarios of solar irradiance and real/reactive power demand profiles, $\Bcal$ describes a suitable $\sigma$-algebra over $\Omega$ and $\prob$ describes the probability measure over $\Omega$. Precisely, define 
\begin{align}
    \v{\alpha}(\omega) \in \Rset^{n-1}, \ \vpd(\omega) \in \Rset^{n-1}, \ \vqd(\omega) \in \Rset^{n-1}
    \label{eq:scenario}
\end{align}
as the profiles of solar irradiance, real power demand and reactive power demand across buses $\ol{\Vcal} := \{2, \ldots, n\}$ in scenario $\omega \in \Omega$, i.e., $\v{\alpha}, \vpd, \vqd$ are $\Bcal$-measurable maps (random variables).
If $\v{\psi} \in \Rset^{n-1}_+$ describes the installed capacities of PV panels across buses in $\ol{\Vcal}$, then the real and reactive power generation from solar panels across the same buses in scenario $\omega$ is given by
\begin{align}
    \vpg(\omega) = \v{\alpha}(\omega) \hadamard \v{\psi}, \quad \vqg(\omega) = \etaG \hadamard \v{\alpha}(\omega) \hadamard \v{\psi},
    \label{scenario.gen}
\end{align}
i.e., $\vpg, \vqg$ are $\Bcal$-measurable for each $\v{\psi}$.
We use the notation $\v{z}_1 \hadamard \v{z}_2 $ to denote the Hadamard (element-wise) product of two arbitrary matrices/vectors $\v{z}_1$ and $\v{z}_2$. Here, $\etaG \in \Rset^{n-1}$ describes a vector of power factors for solar power injection. We remark that today's inverters often operate at unit power factors, for which $\etaG = 0$. The generation profile assumes a direct scaling of the irradiance with the installed capacity--a premise that holds when inverters connecting solar panels track the maximum power point \cite{arab2020maximum}. 
Then, the net real and reactive power injections into buses $\ol{\Vcal}$ are, respectively,
\begin{align}
    \vpinj(\omega) = \vpg(\omega)-\vpd(\omega), \ \vqinj(\omega) =\vqg(\omega)-\vqd(\omega),
    \label{eq:scenario.inj}
\end{align}
where $\vpinj, \vqinj$ are $\Bcal$-measurable.

Derived from Kirchhoff's laws, the power flow equations identify voltages and real/reactive power flows over distribution lines that can sustain these power injections across buses in $\ol{\Vcal}$ in scenario $\omega$. To describe these equations, we introduce additional notation. For the $n-1$ (directed) distribution lines in $\vec{\Ecal}$, let $\v{R} \in \Rset^{n-1}$ and $\v{X} \in \Rset^{n-1}$ collect their resistances and reactances. Define $\v{P}(\omega) \in \Rset^{n-1}$ and $\v{Q}(\omega) \in\Rset^{n-1}$ as the \emph{sending-end} real and reactive power flows in scenario $\omega$ over said lines. Also, let $\v{L}(\omega) \in \Rset^{n-1}$ describe the squared current magnitude over these lines. For the $n$ buses in the distribution network, let $\v{W}(\omega) \in \Rset^{n}$ collect the \emph{squared} voltage magnitudes in scenario $\omega$.

The following matrix operations will prove useful in presenting the power flow equations. For any matrix/vector $\v{z}$, let $\v{z}_+$ compute the positive part of each entry of $\v{z}$. Similarly, define $\v{z}_- := [-\v{z}]_+$. Where unambiguous, we use $\v{z}^2 := \v{z} \hadamard \v{z}$. Finally,  for a vector $\v{z}$ with more than one row, let $\v{\pi}[\v{z}]$ return the same vector with its first row removed.

With this notation, we write the power flow equations as
\begin{subequations}
  \begin{gather}
      \vpinj(\omega) = \v{\pi} \left[ \v{B}_+^\top \v{P}(\omega) - \v{B}_-^\top \left( \v{P}(\omega) - \v{R} \hadamard \v{L}(\omega) \right) \right],
      \label{eq:hcp.flow.real}
      \\
      \vqinj(\omega) = \v{\pi} \left[ \v{B}_+^\top \v{Q}(\omega) - \v{B}_-^\top \left( \v{Q}(\omega) - \v{X} \hadamard \v{L}(\omega) \right) \right],
      \label{eq:hcp.flow.reactive}
      \\
      \v{B}\v{W}(\omega) = 2 \left[\v{R} \hadamard \v{P}(\omega) + \v{X} \hadamard \v{Q}(\omega) \right] 
      \notag
      \\
      \qquad \qquad - (\v{R}^2 + \v{X}^2) \hadamard \v{L}(\omega), 
      \label{eq:hcp.flow.voltage}
      \\
      \left[\v{B}_+ \v{W}(\omega)\right] \hadamard \v{L}(\omega) =  [\v{P}(\omega)]^2 + [\v{Q}(\omega)]^2, \label{eq:hcp.flow.loss}
  \end{gather}
  \label{eq:hcp.flow}
\end{subequations}
This form of the power flow equations for radial distribution networks has been derived in \cite{baran1989optimal}. The first two relations encode power balance at each bus in $\ol{\Vcal}$. The last two equations describe Kirchhoff's voltage law, expressed in terms of squared line currents and real/reactive power flows. 

For a given installed capacity configuration $\v{\psi}$, a scenario $\omega$ defines the real and reactive power injections $\vpinj(\omega), \vqinj(\omega)$ via \eqref{eq:scenario}--\eqref{eq:scenario.inj}. We say that $\v{\Delta}(\omega)$ given by
\begin{align}
    \v{\Delta}(\omega) := \left(\v{P}(\omega), \v{Q}(\omega), \v{L}(\omega), \v{W}(\omega) \right),
\end{align}
\emph{explains} the power injections, when they collectively satisfy \eqref{eq:hcp.flow}. To streamline notation, define the set 
\begin{align}
    \Gcal\left(\vpinj(\omega), \vqinj(\omega)\right) := &\left\{  \v{\Delta}(\omega)   \text{ explains } \vpinj(\omega), \vqinj(\omega) \right\}.
    \label{eq:G.def}
\end{align}
Similarly, we say that a $\Bcal$-measurable $\v{\Delta}$ \emph{explains} an installed capacity configuration $\v{\psi}$, if it explains the induced injection profiles $\prob$-almost surely, i.e.,
\begin{align}
    \v{\Delta} \in \Gcal\left(\v{\alpha}\hadamard\v{\psi} - \vpd, 
   \etaG \hadamard \v{\alpha} \hadamard \v{\psi} - \vqd \right) \ \prob-\text{a.s.}
\end{align}

For a given $\v{\psi}$, a scenario $\omega$ {respects the network constraints}, if it is explained by $\v{\Delta}(\omega)$ that satisfies
\begin{subequations} 
  \begin{gather}
      \ul{\v{W}} \leq \v{W}(\omega) \leq \ol{\v{W}}, 
    \quad
      [\v{P}(\omega)]^2 + [\v{Q}(\omega)]^2 \leq \ol{\v{S}}^2. 
  \end{gather}
  \label{eq:hcp.constraints}
\end{subequations}
The first among these relations define the set of acceptable nodal voltages (typically within 5\% of nominal voltage levels). The second relation imposes thermal limitations on line flows. The inequalities are defined element-wise, where $\ul{\v{W}}, \ol{\v{W}}, \ol{\v{S}}$ are constant vectors of appropriate dimensions. In the risk-sensitive hosting capacity problem, we do not impose such a constraint almost surely, but constrain the risk associated with the statistics of constraint violation. To mathematically describe such a constraint, consider some $\Bcal$-measurable $\v{\Delta}$ that explains the installed capacity configuration $\v{\psi}$. Then, $\v{W} - \ol{\v{W}}$ is a random variable, for which a positive value indicates violation of network constraints. We impose the constraint
\begin{align}
    \CVaR_{\nu}\left[\v{W} - \ol{\v{W}} \right] \leq 0,
    \label{eq:cvar.W.up}
\end{align}
where $\CVaR$ for a random variable is defined in \eqref{eq:cvar_variational}. If $\v{W}$ has a smooth cumulative distribution function, the above constraint imposes that the probability of violating the upper bound on voltage magnitudes is bounded above by $1-\nu$ and the expectation of $\v{W} - \ol{\v{W}}$ over its $\nu$-tail is bounded above by zero.  Thus, such a constraint not only limits the probability of violations but seeks to  control the statistics of violations. $\CVaR$ is translation-invariant, implying that \eqref{eq:cvar.W.up} holds if and only if $ \CVaR_{\nu}\left[\v{W} \right] \leq \ol{\v{W}}$.

Repeating the above exercise for all network constraints in \eqref{eq:hcp.constraints}, we define the following risk-sensitive test for acceptability $ \Ascr(\v{\psi})$ of the configuration $\v{\psi} \in \Rset^{n-1}_+$. We say that $\v{\psi}$ is acceptable, if there exists $\Bcal$-measurable  $\v{\Delta}$, that satisfies
\begin{subequations}
  \begin{align}
   & \v{\Delta} = (\v{P}, \v{Q}, \v{L}, \v{W})  \text{ explains } \v{\psi}, \ \prob-\text{a.s.}, 
    \label{eq:risk.hcp.Delta.explains}
    \\
    & \CVaR_{\nu}\left[\v{W} \right] \leq \ol{\v{W}},
    \label{eq:risk.hcp.constraints.voltage.up}
    \\
    & \CVaR_{\nu}[  - \v{W} ] \leq -\ul{\v{W}}, 
    \label{eq:risk.hcp.constraints.voltage.down}
    \\
    & \CVaR_{\gamma}[ \v{P}^2 + \v{Q}^2 ] \leq \ol{\v{S}}^2,
    \label{eq:risk.hcp.constraints.flow.up}
  \end{align}
  \label{eq:rshcp.const}
\end{subequations}
The formulation can be easily extended to consider different risk levels $\nu$ and $\gamma$ for different buses and distribution lines, to indicate the presence of sensitive equipment. Building on the description of $\Ascr(\v{\psi})$, we next define the risk-sensitive hosting capacity maximization problem $\Oscr$ that seeks to maximize some function of installed capacities $\v{\psi}$ over a set of acceptable subsets of $[0, \ol{\v{\psi}}]$. Here, $\ol{\v{\psi}}$ encodes the maximum surface area available for solar panel installations across buses in $\ol{\Vcal}$.
  \begin{align}
    \underset{\v{\psi} \in [0, \ol{\v{\psi}}], \v{\Delta}}{\text{maximize}} \ \bone^\top \v{\psi}, \ \ 
    \text{subject to} \ \ 
    \eqref{eq:risk.hcp.Delta.explains}, 
    \eqref{eq:risk.hcp.constraints.voltage.up},
    \eqref{eq:risk.hcp.constraints.voltage.down}, 
    \eqref{eq:risk.hcp.constraints.flow.up}.
  \end{align}
  \label{eq:rshcp}
over $\v{\psi} \in \Rset^{n-1}_+$ and $\Bcal$-measurable $\v{\Delta}$. The objective function aims to maximize the total installed capacity of PV panels across the distribution network, where $\bone$ stands for a vector of all ones of appropriate size.

The feasible set of $\v{\Delta}$ for $\Ascr(\v{\psi})$,  and hence for $\Oscr$, is nonconvex, because of \eqref{eq:risk.hcp.Delta.explains}. Specifically, this nonconvexity arises due to the nature of the power flow equations in \eqref{eq:hcp.flow}. For a given power injection $\vpinj(\omega), \vqinj(\omega)$, the set $\Gcal$ defined in \eqref{eq:G.def} is generally nonconvex, owing to the quadratic equality constraint in \eqref{eq:hcp.flow.loss}. It is common practice to \emph{convexify} the power flow model, either via linearization or convex relaxation. For example, the hosting capacity analysis in \cite{taheri2020fast} takes the linearization route. We adopt a second-order cone programming (SOCP)-based relaxation of the set, similar to that in \cite{geng2020probabilistic}, and consider the alternate to \eqref{eq:hcp.flow.loss} for each $\omega$ in
\begin{align}
\left[\v{B}_+ \v{W}(\omega)\right] \hadamard \v{L}(\omega) \geq  [\v{P}(\omega)]^2 + [\v{Q}(\omega)]^2.
\label{eq:hcp.flow.loss.convex}
\end{align}
Then, define the \emph{convex} set
\begin{align}
\begin{aligned}
    \Gcal^\conv(\vpinj(\omega), \vqinj(\omega)) &:= \left\{  \v{\Delta}(\omega) \vert ( \vpinj(\omega), \vqinj(\omega),  \v{\Delta}(\omega))  \right.
    \\
    & \left.  \quad \text{ satisfies } \eqref{eq:hcp.flow.real}, \eqref{eq:hcp.flow.reactive}, \eqref{eq:hcp.flow.voltage}, \eqref{eq:hcp.flow.loss.convex} \right\}.
\end{aligned}
\label{eq:G.conv.def}
\end{align}
The inequality in \eqref{eq:hcp.flow.loss.convex} for each $\omega$ can be written as a collection of second-order cone constraints using the relationship 
\begin{align}
\begin{aligned}
    & z_1 z_2 \geq Z_1^2 +  Z_2^2
    \\
    & \iff \vnorm{\begin{pmatrix} 2Z_1,  2Z_2  , z_1 - z_2\end{pmatrix}^\top}_2 \leq z_1 + z_2
    \\
    & \iff \begin{pmatrix} 2Z_1, 2Z_2, z_1 - z_2, z_1 + z_2\end{pmatrix}^\top \in \KSO
\end{aligned}
\label{eq:SOCP_reform}
\end{align}
for any scalars $Z_1, Z_2, z_1, z_2$, 
where $\vnorm{\cdot}_2$ computes the 2-norm of a vector and $\KSO$, defined as
\begin{align}
    \KSO = \{(\v{y}^\T, \upsilon)^\T \vert \vnorm{\v{y}}_2 \leq \upsilon \},
\end{align}
is the convex second-order cone. The set of $\v{\Delta}$'s that belong to the convex set $\Gcal^\conv$, $\prob$-almost surely, then becomes convex. This convex relaxation of the set of feasible injections defined via Kirchhoff's laws has been widely studied in \cite{bose2014equivalent, farivar2013branch,jabr2006radial}.

The set of $\v{\Delta}(\omega)$ that satisfies the network constraints in \eqref{eq:hcp.constraints} is convex. Thanks to the coherence of the $\CVaR$ measure (see \cite{delbaen2000coherent}), this implies that the set of $\v{\Delta}$ satisfying \eqref{eq:risk.hcp.constraints.voltage.up}, \eqref{eq:risk.hcp.constraints.voltage.down}, \eqref{eq:risk.hcp.constraints.flow.up} is a convex set. 
Using the notation introduced above, we define the test for acceptability $\Ascr^\conv(\v{\psi})$ with a ``relaxed'' power flow model as follows. We say $\v{\psi}$ is acceptable, if there exists $\Bcal$-measurable $\v{\Delta}$, that satisfies  \eqref{eq:risk.hcp.constraints.voltage.up}, 
    \eqref{eq:risk.hcp.constraints.voltage.down}, 
    \eqref{eq:risk.hcp.constraints.flow.up} and
\begin{gather}
    \v{\Delta} \in \Gcal^\conv\left(\v{\alpha}\hadamard\v{\psi} - \vpd, 
    \etaG \hadamard \v{\alpha} \hadamard \v{\psi} - \vqd \right) \ \prob-\text{a.s.} 
    \label{eq:risk.hcp.Delta.explains.conv}
\end{gather}
The corresponding convexified hosting capacity maximization problem $\Oscr^\conv$ over $\v{\psi} \in \Rset^{n-1}_+$ and $\Bcal$-measurable $\v{\Delta}$  becomes 
  \begin{align}
    \underset{\v{\psi} \in [0, \ol{\v{\psi}}], \v{\Delta}}{\text{maximize}} \ \bone^\top \v{\psi}, \ \ 
    \text{subject to} \ \ 
    \eqref{eq:risk.hcp.Delta.explains.conv}, 
    \eqref{eq:risk.hcp.constraints.voltage.up},
    \eqref{eq:risk.hcp.constraints.voltage.down}, 
    \eqref{eq:risk.hcp.constraints.flow.up}.
  \end{align}
  \label{eq:rshcp.conv}

\section{A Scenario-Based Approach to Solve \texorpdfstring{$\Oscr^\conv$}{Oc}}
\label{sec:scenario}

The set of $\v{\Delta}$'s satisfying $\Ascr^\conv(\v{\psi})$ and the set of $\v{\psi}, \v{\Delta}$ satisfying $\Oscr^\conv$ are convex. However, finding/optimizing $\Bcal$-measurable  $\v{\Delta}$ can be challenging. If $\Omega$ is compactly supported, $\v{\Delta}$'s are infinite-dimensional functions defined over a compact set. To circumvent the difficulty of optimizing over function spaces, we take a \emph{sampling} approach to solve $\Oscr^\conv$.


Consider $K$ independent and identically distributed (i.i.d.) samples $\omega^1, \ldots, \omega^K$ from $\Omega$, drawn according to $\prob$. With $K$ samples of solar irradiance, real and reactive power demands across buses in $\ol{\Vcal}$, denoted by the vector tuples
$\{ (\v{\alpha}^k, \vpd^k, \vqd^k) \}_{k = 1}^K$, consider 
\begin{align}
    \v{\Delta}(\omega^k) = \v{\Delta}^k = (\v{P}^k, \v{Q}^k, \v{L}^k, \v{W}^k).
\end{align}
The sampled variant of \eqref{eq:risk.hcp.Delta.explains.conv} becomes 
\begin{align}
    \v{\Delta}^k \in \Gcal^\conv\left(\v{\alpha}^k\hadamard\v{\psi} - \vpd^k, 
    \etaG \hadamard \v{\alpha}^k \hadamard \v{\psi} - \vqd^k \right),
\end{align}
for $k=1,\ldots,K$.
Leveraging the variational form of $\CVaR$ from \eqref{eq:cvar_variational}, we write the scenario-based counterpart of \eqref{eq:risk.hcp.constraints.voltage.up} as
\begin{align}
\begin{aligned}
    & \min_{\ol{\v{w}}} \ \left\{ \ol{\v{w}} + \frac{1}{1-\nu}\frac{1}{K}\sum_{k=1}^K 
    \left[ \v{W}^k - \ol{\v{w}} \right]_+ \right\} \leq \ol{\v{W}}
    \\
    & \iff \begin{cases}
    \ol{\v{w}} + \frac{1}{1-\nu}\frac{1}{K}\sum_{k=1}^K \ol{\v{t}}^k_w \leq \ol{\v{W}},
    \\
    \ol{\v{t}}^k_w \geq \v{W}^k - \ol{\v{w}}, \ \ol{\v{t}}^k_w \geq 0,
    \end{cases}
\end{aligned}
\label{eq:cvar.reform}
\end{align}
for {some} $\ol{\v{w}} \in \Rset^{n}, \ol{\v{t}}^k_w \in \Rset^{n}$, $k=1,\ldots,K$. Thus, 
one can drop the sampled $\CVaR$ constraint, and instead include the 
latter pair of deterministic constraints listed above, and optimize over $\ol{\v{w}}, \ol{\v{t}}^k_w$, $k=1,\ldots,K$. We repeat this procedure for \eqref{eq:risk.hcp.constraints.voltage.down} and \eqref{eq:risk.hcp.constraints.flow.up} and present the sampled counterpart 
$\Oscr^\conv_{\sample}$ of $\Oscr^\conv$.
\begin{subequations} %
\begin{alignat}{2} %
    & {\text{maximize}} \quad \bone^\top \v{\psi},
    \notag
    \\
    & \text{subject to} 
    \notag
    \\
    &
    \ 0\leq \v{\psi}\leq \ol{\v{\psi}},
    \label{eq:rshcp.box}
    \\
    & 
    \ \v{\alpha}^k \hadamard \v{\psi} - \vpd^k
    = \v{\pi} \left[ \v{B}_+^\top \v{P}^k - \v{B}_-^\top \left( \v{P}^k - \v{R} \hadamard \v{L}^k \right) \right],
    \label{eq:rshcp.net.1}
    \\
    & 
    \ \etaG \hadamard \v{\alpha}^k \hadamard \v{\psi} - \vqd^k
    = \v{\pi} \left[ \v{B}_+^\top \v{Q}^k - \v{B}_-^\top \left( \v{Q}^k - \v{X} \hadamard \v{L}^k \right) \right],
    \label{eq:rshcp.net.2}
    \\
    &
    \ \v{B}\v{W}^k = 2 \left[\v{R} \hadamard \v{P}^k + \v{X} \hadamard \v{Q}^k \right] 
    - (\v{R}^2 + \v{X}^2) \hadamard \v{L}^k,
    \label{eq:rshcp.net.3}
    \\
    &
    \ \left[\v{B}_+ \v{W}^k\right] \hadamard \v{L}^k \geq  [\v{P}^k]^2 + [\v{Q}^k]^2,
    \label{eq:rshcp.net.4}
    \\
    & 
    \ \ol{\v{w}} + \frac{1}{1 - \nu} \frac{1}{K}\sum_{k = 1}^K \ol{\v{t}}^k_w \le \ol{\v{W}},
    \ \ol{\v{t}}^k_w \ge \v{W}^k - \ol{\v{w}},
    \label{eq:rshcp.cvar.1}
    \\
    & 
    \ \ul{\v{w}} + \frac{1}{1 - \nu} \frac{1}{K}\sum_{k = 1}^K \ul{\v{t}}^k_w \leq -\ul{\v{W}},
    \ \ul{\v{t}}^k_w \geq - \v{W}^k - \ul{\v{w}},
    \label{eq:rshcp.cvar.2}
    \\
    & 
    \ \v{s} + \frac{1}{1 - \gamma}\frac{1}{K} \sum_{k = 1}^K \v{t}^k_s \leq \ol{\v{S}}^2,
    \label{eq:rshcp.cvar.3}
    \\
    &
    \ \v{t}^k_s \ge [\v{P}^k]^2 + [\v{Q}^k]^2 - \v{s},
    \label{eq:rshcp.cvar.3.2}\\
    &
    \ \ol{\v{t}}^k_w\geq \v{0},\,\ul{\v{t}}^k_w\geq \v{0},\,\v{t}^k_s \ge \v{0}, \ \text{for } k = 1, \dots, K
    \label{eq:rshcp.nonneg}
\end{alignat} %
\label{eq:rshcp.saa} %
\end{subequations} %
over the variables $\v{\psi}, \v{P}^k, \v{Q}^k, \v{W}^k, \v{L}^k, \ol{\v{w}}, \ul{\v{w}}, \v{s}, \ol{\v{t}}^k_w, \ul{\v{t}}^k_w, \v{t}^k_s$. All constraints in $\Oscr^\conv_\sample$ are linear, except those in \eqref{eq:rshcp.net.4} and \eqref{eq:rshcp.cvar.3.2}. One can utilize \eqref{eq:SOCP_reform} to cast them as second-order cone constraints. As a result, $\Oscr^\conv_\sample$ becomes an SOCP--a \emph{convex} optimization problem. Such SOCP's can be solved using off-the-shelf solvers that employ interior-point algorithms. 

The \CVaR{}-based formulation of $\Oscr^\conv$ and its scenario-based variant in $\Oscr^\conv_\sample$  provide a structured framework to identify optimal installation configurations. The statistics of network constraint violation across scenarios impact the choice of the optimal installation profile. 
While our approach shares many parallels with the BayesOpt framework with probabilistic constraints in \cite{geng2020probabilistic}, there are important differences. To explain the difference, consider $\Oscr^\conv$ with probabilistic constraint enforcement, i.e., replace $\CVaR_\nu[\v{W}] \leq \ol{\v{W}}$ with $\prob\{ \v{W} - \ol{\v{W}} >0\} \leq 1-\nu$. With scenarios, the probabilistic constraint amounts to imposing
\begin{equation}\label{eq:SAA_chance}
    \frac{1}{K}\sum_{k=1}^K\Iset_{\{\v{W}^k - \ol{\v{W}} >0 \}}\leq 1-\nu,
\end{equation}
where $\Iset$ is the indicator function that evaluates to unity when its argument is true, and becomes zero otherwise. Tackling such constraints with indicator functions within our framework requires the solution of a nonconvex mixed-integer SOCP, which is typically more computationally challenging than solving the SOCP problem $\Oscr^\conv_\sample$ with \CVaR{}-based constraints. Note that a solver for $\Oscr^\conv_\sample$ will typically update all optimization variables together towards an optimal solution. In contrast, the probability of constraint violation in the left-hand-side of \eqref{eq:SAA_chance} must be computed, each time the candidate installation profile $\v{\psi}$ changes in search of an optimal solution. More precisely, as $\v{\psi}$ is updated to $\v{\psi}'$, the already-computed $\v{\Delta}^k$ for scenario $\omega^k$ with installation profile $\v{\psi}$ is discarded, and must be recomputed from scratch with an installation profile $\v{\psi}' \neq \v{\psi}$. Lacking a convex structure for the set of $\v{\Delta}^k$'s that satisfy \eqref{eq:SAA_chance}, it becomes difficult to recycle the computation of $\v{\Delta}^k$ \emph{across} candidate installation profiles. Our formulation does not explicitly evaluate \CVaR{} over scenarios for each $\v{\psi}$. Rather, we leverage the variational characterization of \CVaR{} in \eqref{eq:cvar_variational} to formulate $\Oscr^\conv_\sample$ over $\v{\psi}$, $\v{\Delta}^k$'s and auxiliary variables, where a solver can exploit the convexity of the problem to iteratively update these variables together.

The lack of convexity in the left-hand-side of \eqref{eq:SAA_chance} makes it challenging to find natural directions to update $\v{\psi}$. The Bayesian optimization framework in \cite{geng2020probabilistic} tackles this challenge through a combination of two steps. First, the constraint in \eqref{eq:SAA_chance} is included through a penalty in the objective function to construct an unconstrained optimization problem. Then, the resulting objective is minimized via kernel smoothing of the objective function. Running such algorithms requires careful choice of the penalty factor and the kernel. Our \CVaR{}-based formulation being convex, avoids such manual tuning.

Our formulation of $\Oscr^\conv$ and $\Oscr^\conv_\sample$ utilizes the SOCP-based relaxation of the set of feasible injections, characterized by Kirchhoff's laws. The setup is flexible in that one can utilize any ``convexified'' power flow model to obtain a convex programming formulation of the \CVaR{}-sensitive hosting capacity problem. This approach offers a sharp contrast to those that rely heavily on the choice of a particular power flow model. For example, the sample reduction technique in \cite{taheri2020fast}  crucially depends on a linear power flow model, where constraint violations are processed through a quadratic penalty and optimized, permitting the use of multi-parametric quadratic programming theory to assess the feasibility of an installation configuration $\v{\psi}$. Such a method does not naturally generalize to SOCP-based power flow models.

\subsection{Numerical Experiments with   \texorpdfstring{$\Oscr^\conv_\sample$}{Oc}}
\label{sec:numerics}

We solved $\Oscr^\conv_\sample$ on two networks: a 3-bus network for illustrative purposes and the Southern California Edison's 56-bus network from \cite{gan2014exact}, studied previously in \cite{geng2020probabilistic}, to demonstrate scalability. Simulations on the 3-bus network were performed on an i5-4690k machine with 16GB RAM, while the 56-bus simulations were performed on an AWS m4.10xlarge instance. All results were generated using MOSEK version 9.3. Load data and DER energy profiles were taken at 10 minute intervals over a 365-day period (52.56K intervals) using data from Southern California Edison.
For all experiments, DER installation power factors were set to $0.97$ ($\eta = 0.251$), and the feeder voltage was fixed at $w_1 = 1$ p.u. Data and code are available on GitHub at {\url{https://github.com/amadavan/RSHC}}.

\subsubsection{On a three-bus network} 
\begin{figure}[ht]
    \centering
    \subfloat[\label{fig:3bus.time}]{
     \includegraphics[width=0.49\linewidth]{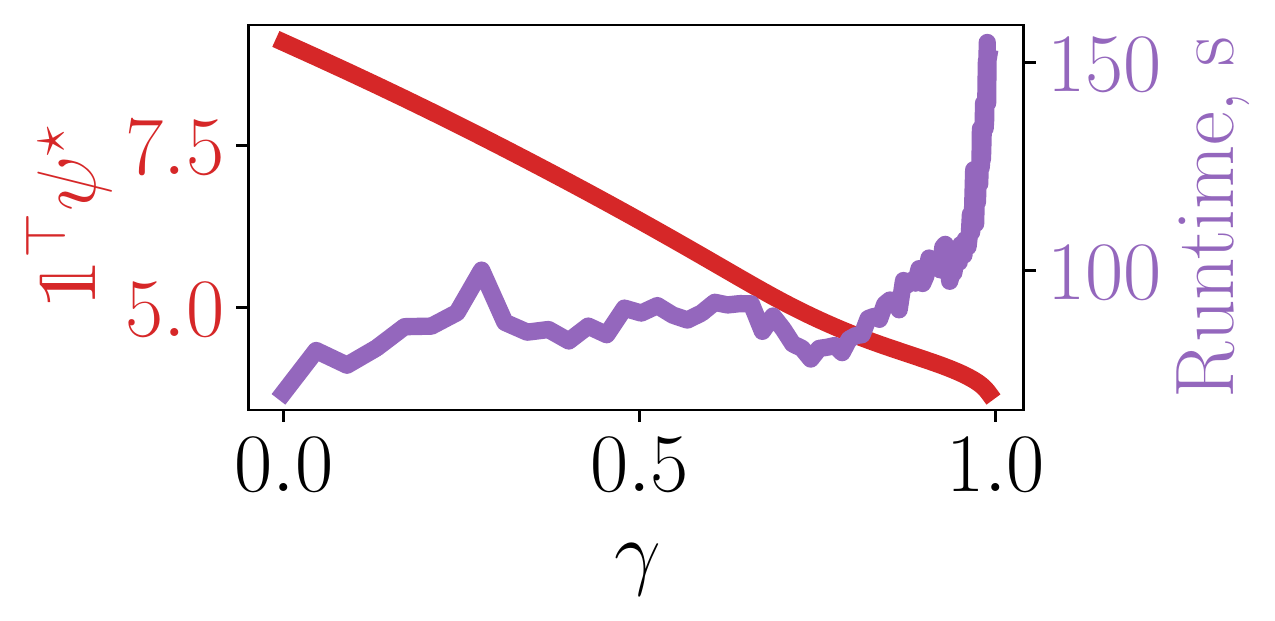}
    }
    \subfloat[\label{fig:3bus.ssa}]{
      \includegraphics[width=0.47\linewidth]{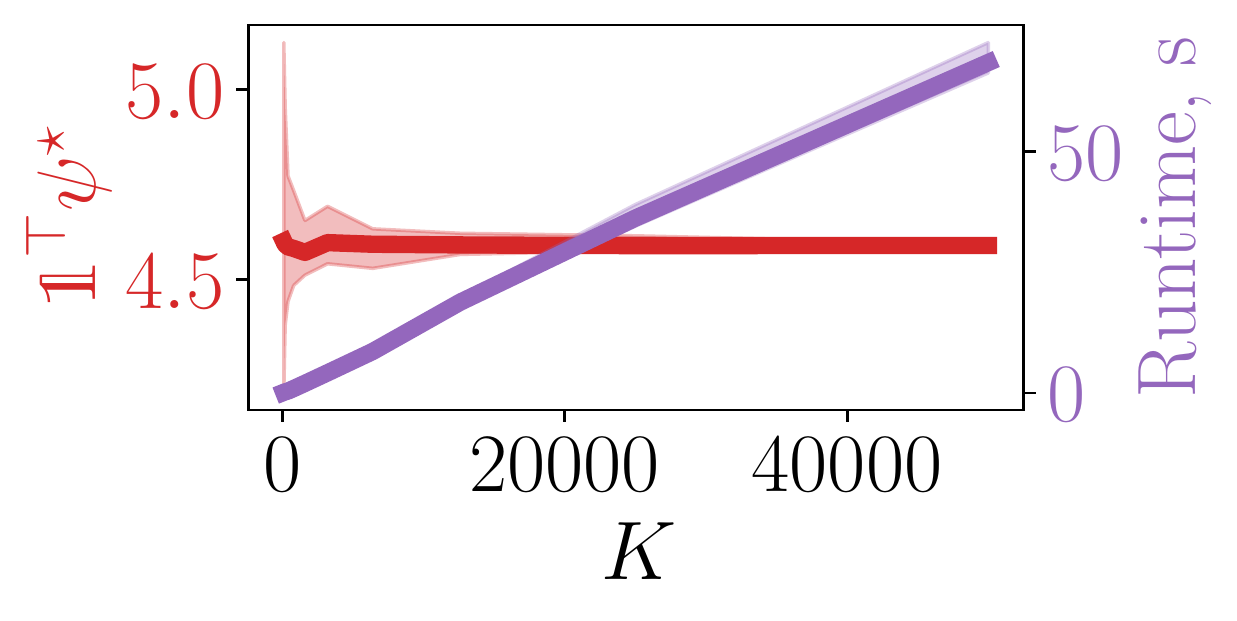}
    }
    \caption{Experiments on the 3-bus network, showing \protect\subref{fig:3bus.time} the effect of the \CVaR{} parameter $\gamma$ on the maximum capacity and runtime, and \protect\subref{fig:3bus.ssa} the deviation in solutions with increasing samples with $\gamma = \nu = 0.8$.}
    \label{fig:3bus}
\end{figure}

In Figure \ref{fig:3bus.time}, we examine the effect of the risk parameters ($\gamma=\nu$) on both the solution and the run-time. The optimal aggregate capacity $\bone^\top \v{\psi}^\star$ decreases with risk aversion. This is anticipated as a higher risk parameter leads to a tighter risk-sensitive constraint, shrinking the feasible set. In turn, this leads to a decreased maximum total installed capacity. We believe that increasing run-times are a consequence of the tightening constraints.




The problem dimension scales linearly with the number of scenarios $K$. Each scenario introduces $9n - 5$ constraints and $7n - 4$ variables, where $n$ is the number of buses. In Figure \ref{fig:3bus.ssa}, we evaluate the effect of sub-sampling on the resulting solution, plotting the mean optimal solution with deviations over 20 runs per sample size. Fewer samples result in greater variance in the solution, but reduced run-times.

\subsubsection{On a 56-bus distribution network}
\begin{figure}[!ht]
    \centering
    \subfloat[\label{fig:56bus.time}]{
     \includegraphics[width=0.48\linewidth]{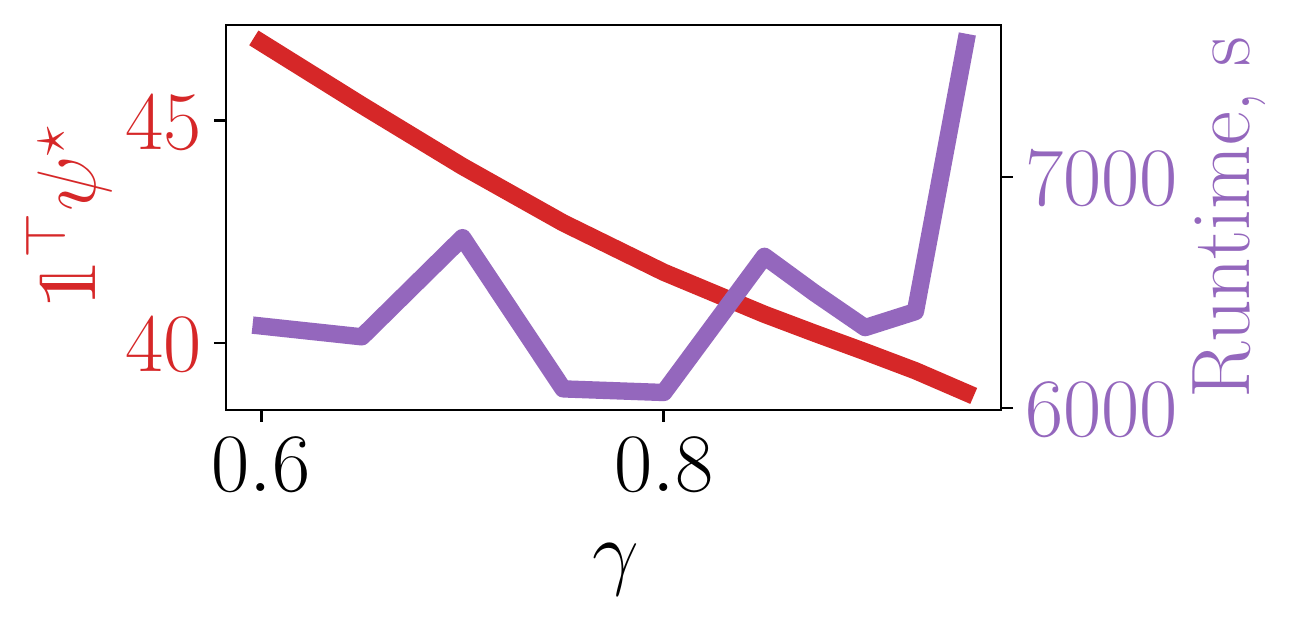}
    }
    \subfloat[\label{fig:56bus.ssa}]{
      \includegraphics[width=0.49\linewidth]{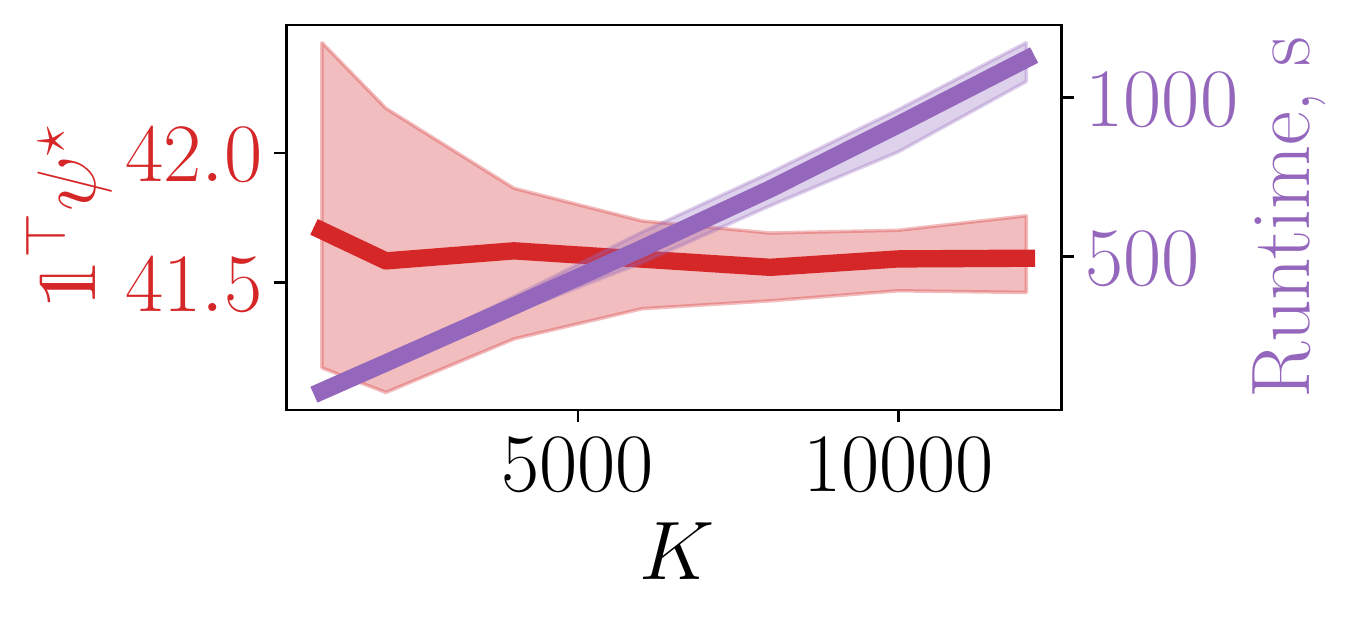}
    }
    \caption{Experiments on the 56-bus network with $\nu = 0.9$, showing \protect\subref{fig:56bus.time} the effect of the \CVaR{} parameter $\gamma$ on the maximum capacity and runtime, and \protect\subref{fig:56bus.ssa} the deviation in solutions with increasing samples with $\gamma = 0.8$.}
    \label{fig:rshcp.56bus}
\end{figure}

Figure \ref{fig:rshcp.56bus} shows the effect of the risk parameter $\gamma$ (keeping $\nu$ constant) on both run-times and optimal objective values attained with $K=$52.56K scenarios. As in the 3-bus example, the objective value decreases with $\nu$. The algorithm run-times are 1-2 hours for each value of risk parameter considered, and increase significantly for $\gamma$ close to $1$.

\begin{figure}[!h]
    \centering
    \includegraphics[width=0.8\linewidth]{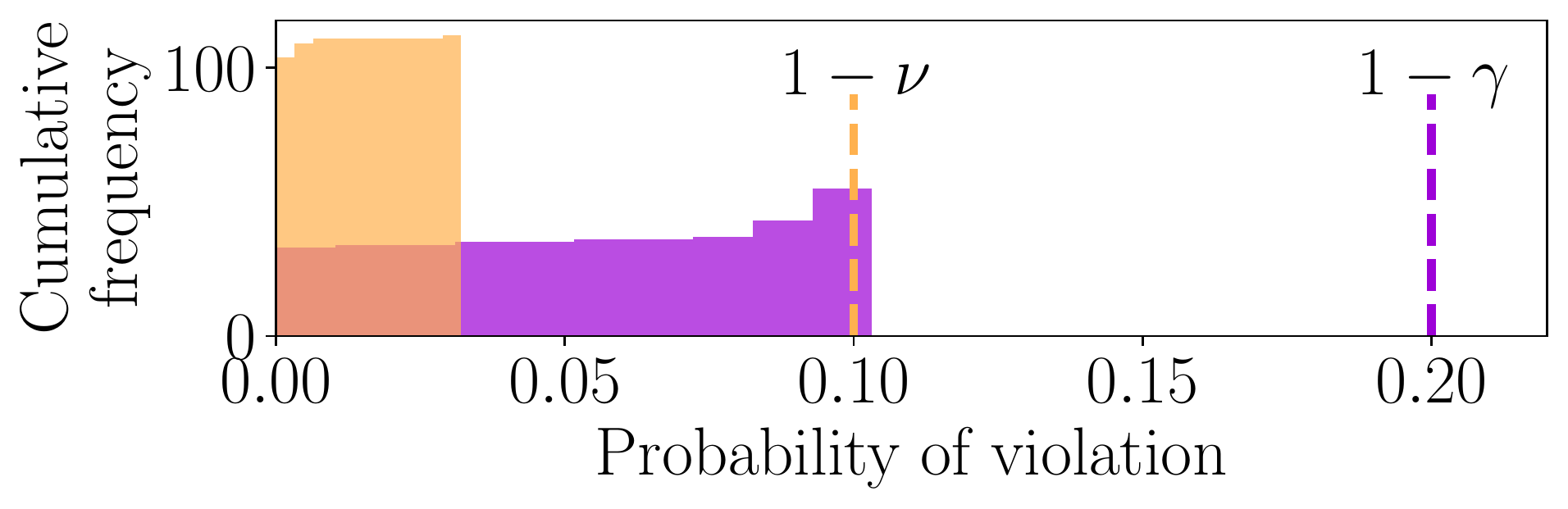}
    \caption{Cumulative frequency histogram of proportions of 52,560 scenarios over which upper/lower voltage constraints across buses and power flow constraints across distribution lines for the 56-bus network are violated with $\nu = 0.9$,  $\gamma = 0.8$.}
    \label{fig:56bus.vio}
\end{figure}

As previously noted, for a random variable $\zeta$, the constraint $\CVaR{}_{\delta}[{\zeta}] \le 0$ is a convex inner approximation of the constraint $\prob[\zeta \ge 0] \le 1 - \delta$. In other words, a configuration $\v{\psi}$ for which the CVaR of voltage variations and power flows satisfies the constraints in $\Oscr^\conv$, they must satisfy the corresponding probabilistic constraints. To illustrate the same empirically with the optimal solution $\v{\psi}^\star$, consider the solution obtained with $\nu = 0.9$ and $\gamma = 0.8$. For this solution, we evaluate the proportion of the 52.56K $\v{\Delta}^k$'s for which each of the 110 upper and lower voltage limits and 55 power flow limits are violated. Figure \ref{fig:56bus.vio} shows the cumulative distribution histogram of the proportions of voltage violations across the buses and power flow violations across the lines. The support of voltage violation proportions lie fully to the left of $1-\nu$ and the power flow violation proportions to the left of $1-\gamma$. That is, our \CVaR{}-based constraint enforcement automatically guarantees probabilistic constraint enforcement.


While our reported solutions used MOSEK, we have compared our results with an implementation in Gurobi. For high risk-aversion parameters, e.g., $\nu = \gamma = 0.9$, Gurobi took almost twice as long to produce an optimal solution, compared to MOSEK, even with small number of scenarios. However, in performing our simulations, MOSEK tended to yield a stable optimal solution ($\approx$ the same result as Gurobi) the fastest, when the objective function was scaled by a large parameter $\sim 10^{9}$. Run-times reduced by as much as by a factor of two with such scaling. The Gurobi implementation remained immune to objective function scaling.


Compared to the BayesOpt algorithm in \cite{geng2020probabilistic}, our implementation is \emph{centralized}, and naturally requires more memory than a solution architecture that solves scenario-wise power flow problems. A side-by-side comparison with earlier works  remains difficult, given that optimization with probabilistic constraint enforcement  has no optimality guarantees and its outcome depends strongly on implementation details. 

One might surmise that our formulation of $\Oscr^\conv_\sample$ can be parallelized using an algorithmic architecture such as generalized Benders' decomposition in \cite{geoffrion1972generalized}. Indeed, one can devise such a method that processes scenarios $\omega^k$'s in parallel. However, our implementation of generalized Benders' decomposition for $\Oscr^\conv_\sample$ led to increasingly larger master problems, resulting from feasibility cuts from individual scenarios, that proved computationally prohibitive. We abandoned that route, given that the centralized implementation ran in reasonable time. A more careful implementation with lazy constraint generation protocols is left for future endeavors.

\section{An Incremental Algorithm to Solve \texorpdfstring{$\Ascr^\conv(\v{\psi})$}{Ac} with Scenarios}
\label{sec:feasibility}

\newcommand{\inS}{\textrm{in}}
\newcommand{\outS}{\textrm{out}}

Setting aside the issue of optimality, we now study a scenario-based approach to solve $\Ascr^\conv(\v{\psi})$ and certify whether the risk of network constraint violations associated with certain configuration of solar installation capacities $\v{\psi} \in \Rset^{n-1}_+$ is acceptable. Recall that $\v{\psi}$ is deemed acceptable, if there exists a $\Bcal$-measurbable $\v{\Delta}$ that satisfies \eqref{eq:risk.hcp.constraints.voltage.up}, 
\eqref{eq:risk.hcp.constraints.voltage.down}, 
\eqref{eq:risk.hcp.constraints.flow.up} and \eqref{eq:risk.hcp.Delta.explains.conv}. 
Similar to the design of $\Oscr^\conv_\sample$, one can certify the acceptability of $\v{\psi}$ with a list of $K$ i.i.d. samples $\omega^1, \ldots, \omega^K$ from $\Omega$ via the convex optimization problem $\Ascr^\conv_\sample(\v{\psi})$, described by
\begin{align}
    \Ascr^{\conv,\star}_\sample(\v{\psi}) := {\text{maximize}} \  0,
    \  \text{subject to }  \ \eqref{eq:rshcp.net.1}-\eqref{eq:rshcp.nonneg},
    \label{eq:feasibility}
\end{align}
over $\v{P}^k, \v{Q}^k, \v{W}^k, \v{L}^k$,   $\ol{\v{w}}, \ul{\v{w}}, \v{s}, \ol{\v{t}}^k_w, \ul{\v{t}}^k_w, \v{t}^k_s$ for $k=1,\ldots,K$. 
If \eqref{eq:feasibility} admits a feasible point, then $\Acal^\star(\v{\psi}) = 0$ and $\v{\psi}$ is acceptable. Otherwise, $\Acal^\star(\v{\psi})=-\infty$ and $\v{\psi}$ is unacceptable.

\begin{figure}[!ht]
  \centering
  \includegraphics[width=0.46\linewidth]{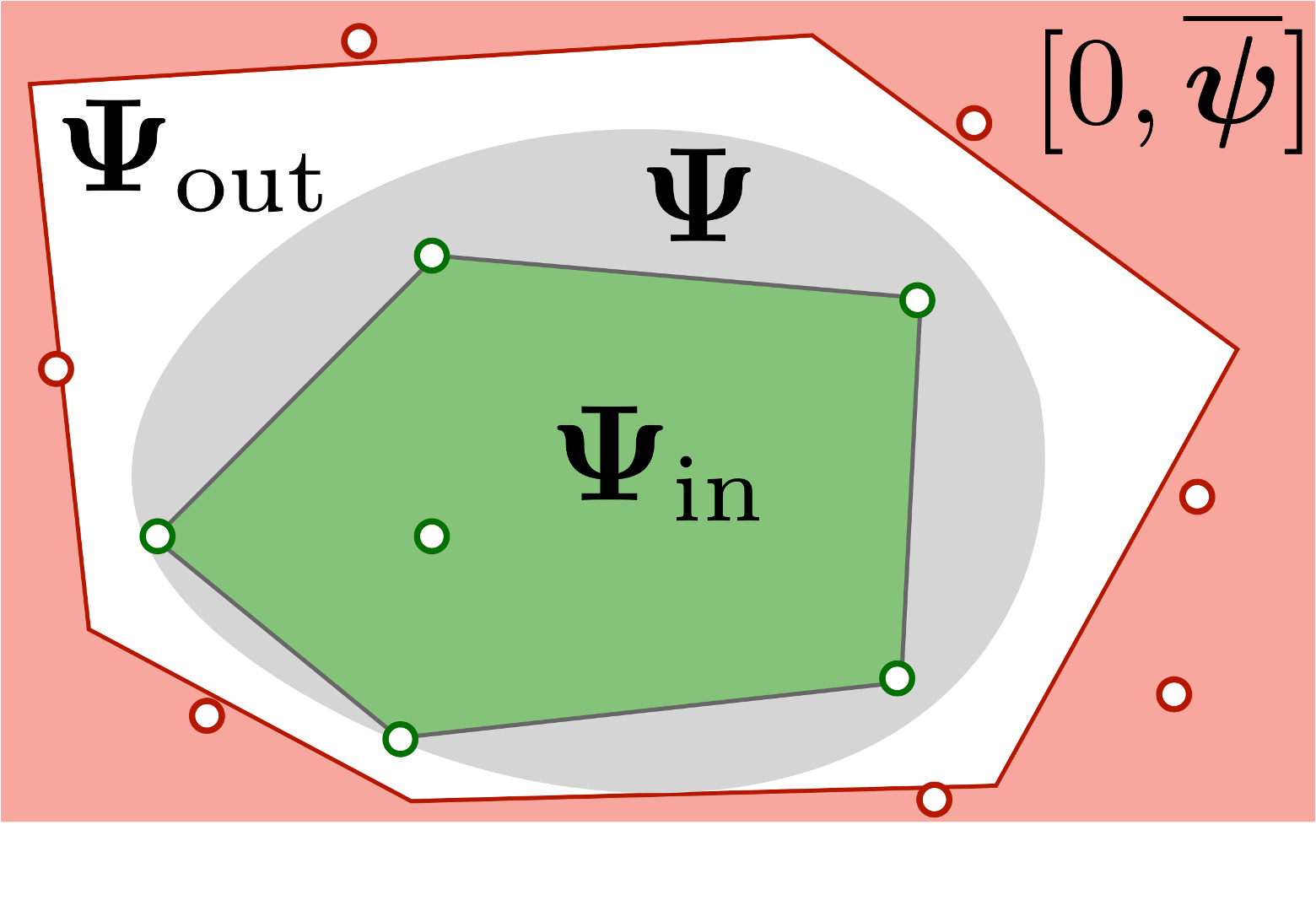}
  \caption{Inner (green) and outer (white) sets portray $\v{\Psi}_\inS$ and $\v{\Psi}_\outS$, respectively, that approximate the set of acceptable installations $\v{\Psi}$ (gray). The outer frame encodes the set $[0, \ol{\v{\psi}}]$. The sets $\Aset_\psi$ and $\Uset_\psi$ comprise the points with green and red borders, respectively.}
  \label{fig:innerouter}. 
\end{figure}
One can always solve \eqref{eq:feasibility} to certify whether $\v{\psi}$ is acceptable. This problem has a similar time-complexity as $\Oscr^\conv_\sample$. In the sequel, we develop an \emph{incremental algorithm} to certify acceptability. That is, we first attempt to leverage the knowledge of a list of acceptable and unacceptable configurations to certify whether a new candidate configuration is acceptable. Our design is such that this check is often much faster than solving \eqref{eq:feasibility}. If unsuccessful, we then solve \eqref{eq:feasibility} and use the new configuration to update the knowledge about  acceptable and unacceptable configurations. As will be evident, such a method drastically reduces the instances of $\v{\psi}$'s for which \eqref{eq:feasibility} must be solved to certify acceptability with more tests. Thus, the more we test, the less we require to solve \eqref{eq:feasibility}. Such a testing paradigm is useful, especially to test acceptability of installation configurations that are close to each other, e.g., those that lie on likely solar adoption paths.

Let $\v{\Psi}$ denote the set of all acceptable configurations of physically realizable installed solar capacities, i.e., $\v{\Psi} \subseteq [0, \ol{\v{\psi}}]$. Thanks to the convexity of CVaR, $\v{\Psi}$ is a {convex set}, shaded in gray in Figure \ref{fig:innerouter}. We maintain polyhedral inner and outer approximations $\v{\Psi}_\inS$ and  $\v{\Psi}_\outS$, respectively, of $\v{\Psi}$ satisfying
\begin{align}
    \v{\Psi}_\inS \subseteq \v{\Psi} \subseteq \v{\Psi}_\outS \subseteq [0, \ol{\v{\psi}}].
\end{align}
These sets are visualized in Figure \ref{fig:innerouter}.
We construct $\v{\Psi}_\inS$ and $\v{\Psi}_\outS$ incrementally from prior acceptability tests. For a new candidate $\v{\psi}$, if $\v{\psi} \in \v{\Psi}_\inS$, then $\v{\psi}$ is acceptable. On the other hand, if $\v{\psi} \notin \v{\Psi}_\outS$, then $\v{\psi}$ is unacceptable. If neither of these tests certify the acceptability of $\v{\psi}$, then we run \eqref{eq:feasibility}. The result of the optimization program either grows $\v{\Psi}_\inS$ or shrinks $\v{\Psi}_\outS$, thus expanding the space of $\v{\psi}$'s, where acceptability can be certified without running \eqref{eq:feasibility}.
In the rest of this section, we outline the incremental construction of $\v{\Psi}_\inS, \v{\Psi}_\outS$ and how we test $\v{\psi} \in \v{\Psi}_\inS$ or $\v{\psi} \notin \v{\Psi}_\outS$.
The method for testing acceptability of $\v{\psi}$ is summarized in Algorithm \ref{alg:rsasct}.


With a list  $\Aset_\psi := 
\{\v{\psi}^a[1], \ldots, \v{\psi}^a[M]\}$ of acceptable configurations, we construct $\v{\Psi}_\inS$ as their \emph{convex hull}, 
\begin{gather}
\begin{gathered}
\v{\Psi}_\inS  = \textrm{conv}(\Aset_\psi) = \left\{ \sum_{i=1}^M\beta_i\v{\psi}^a[i]  :  \v{\beta} \in \Rset^M_+, \bone^\top \v{\beta} = 1 \right\}.
\end{gathered}
\label{eq:def.psi.in}
\end{gather}
Testing if $\v{\psi} \in \v{\Psi}_\inS$ can be solved as a linear program over $\v{\beta}\in\Rset^{M}$. If $\v{\psi}$ is certified acceptable from running \eqref{eq:feasibility}, then $\Aset_\psi \gets \Aset_\psi \cup \v{\psi}$ and define again $\v{\Psi}_\inS = \textrm{conv}(\Aset_\psi)$. We grow $\v{\Psi}_\inS$ incrementally, starting from the null set.


Next, we describe how we construct $\v{\Psi}_\outS$, starting with $[0, \ol{\v{\psi}}]$.  In the event that \eqref{eq:feasibility} finds $\v{\psi}$ unacceptable, we construct a \emph{feasibility cut}, borrowing the idea from generalized Benders' decomposition in \cite{geoffrion1972generalized}. To explain cut generation, define $\v{x}$ as the vectorized concatenation of all variables in \eqref{eq:feasibility} and  write \eqref{eq:feasibility} as
\begin{subequations}
\begin{alignat}{2}
    \hspace{-0.1in}
    \Ascr^{\conv,\star}_\sample(\v{\psi}) = \,& \underset{\v{x}}{\text{maximize}} & \quad & 0,
    \notag
    \\
    \label{eq:socp.psi_const}& 
    \text{subject to} && 
    \v{C}_\psi \v{\psi} = \v{D}_\psi \v{x} + \v{E}_\psi,
    \\
    \label{eq:socp.lin_ineq_const}
    &&& \v{D}\v{x} \leq \v{E},
    \\
    \label{eq:socp.cone_const}
    &&&\begin{pmatrix}\v{F}_{i}\v{x}  + \v{\v{G}}_{i}
    \\
    \v{f}^{\top}_{i}\v{x} + g_{i}\end{pmatrix}\in \KSO,
    \\
    &&& i=1,\ldots,I. \notag
\end{alignat}
\label{eq:socp}
\end{subequations}
The equalities \eqref{eq:rshcp.net.1} and \eqref{eq:rshcp.net.2} are written compactly as \eqref{eq:socp.psi_const}. 
The equality in \eqref{eq:rshcp.net.3} can be written as two inequalities. These, together with 
\eqref{eq:rshcp.cvar.1}--\eqref{eq:rshcp.cvar.3} and \eqref{eq:rshcp.nonneg} become examples of  \eqref{eq:socp.lin_ineq_const}. The nonlinear constraints in \eqref{eq:rshcp.net.4} and \eqref{eq:rshcp.cvar.3.2} are written as $I$ SO cone constraints \eqref{eq:socp.cone_const} using \eqref{eq:SOCP_reform}.
To lighten notation, assume $I = 1$ and consider the following Lagrangian function
\begin{align}
\begin{aligned}
    &\Lcal(\v{x}, \v{\lambda}, \v{\mu}, \v{\mu}_1, \mu_2) 
    \\
    &
    = \v{\lambda}^{\top} ( \v{C}_\psi \v{\psi} - \v{D}_\psi \v{x} - \v{E}_\psi )  
    + \v{\mu}^{\top} ( \v{D}\v{x} - \v{E} ) \\
    &\qquad - \v{\mu}_1^{\top} ( \v{F}\v{x} + \v{G} )  - \mu_2 ( \v{f}^{\top}\v{x} + g )
\end{aligned}
\end{align}
for Lagrange multipliers $\v{\lambda}, \v{\mu}, \v{\mu}_1, \mu_2$ of compatible dimensions. Then, \eqref{eq:socp} can be written as
\begin{equation}
    -\Ascr^{\conv,\star}_\sample(\v{\psi}) = \underset{\v{x}}{\text{min}} \  \underset{\substack{\v{\lambda}, \v{\mu} \ge 0, \\ (\v{\mu}_1, \mu_2) \in \KSO}} {\text{max}} \ \Lcal(\v{x}, \v{\lambda}, \v{\mu}, \v{\mu}_1, \mu_2).
\end{equation}
The above representation leverages the fact that $\KSO$ is a self-dual convex cone (see  \cite{boyd2004convex} for details). Weak duality implies
\begin{equation}
    -\Ascr^{\conv,\star}_\sample(\v{\psi}) \ge \underset{\substack{\v{\lambda}, \v{\mu} \ge 0, \\ (\v{\mu}_1, \mu_2) \in \KSO}}{\text{max}} \ \ \underset{\v{x}}{\text{min}} \ \ \Lcal(\v{x}, \v{\lambda}, \v{\mu}, \v{\mu}_1, \mu_2),
\end{equation}
where the right hand side of the above inequality equals the optimal cost of the dual problem. Solving the inner minimization over $\v{x}$ in the above relation, yields
\begin{equation}
\begin{aligned}
  -\Ascr^{\conv,\star}_\sample(\v{\psi}) \ge\,\,& \underset{\v{\lambda}, \v{\mu}, \v{\mu}_1, \mu_2}{\text{maximum}} && ( \v{C}_\psi \v{\psi} - \v{E}_\psi )^\top \v{\lambda} 
  \\
  &&&\quad 
  - \v{E}^{\top} \v{\mu} - \v{G}^{\top} \v{\mu}_1 - \mu_2 g, \\
  & \text{subject to} && \v{D}^{\top} \v{\mu} = \v{D}_{\psi}^{\top} \v{\lambda} + \v{F}^{\top} \v{\mu}_1 + \v{f} \mu_2, \\
    &&& \v{\mu} \ge 0, \, (\v{\mu}_1^{\top}, \mu_2)^\top \in \KSO.
\end{aligned}
\label{eq:socp.dual}
\end{equation}
The dual program for \eqref{eq:socp} is an SOCP, for which the origin is always feasible. The feasible set for this dual problem is a convex cone.\footnote{$\Scal$ is a convex cone, if $s_1, s_2 \in\Scal,  \kappa_1, \kappa_2 \geq 0$ implies $\kappa_1 s_1 + \kappa_2 s_2 \in \Scal$.} Call the objective function of this problem as $\Dcal(\v{\lambda}, \v{\mu}, \v{\mu}_1, \mu_2; \v{\psi})$. Then, for any feasible $\v{\lambda}^\circ, \v{\mu}^\circ, \v{\mu}_1^\circ, \mu_2^\circ$ yielding a positive dual objective, we have
\begin{align}
\begin{aligned}
    &\Dcal(\v{\lambda}^\circ,\v{\mu}^\circ,\v{\mu}_1^\circ,\mu_2^\circ; \v{\psi}) > 0 
    \\
    &\implies
    \lim_{\kappa \to \infty} \Dcal(\kappa \v{\lambda}^\circ,\kappa \v{\mu}^\circ, \kappa \v{\mu}_1^\circ, \kappa \mu_2^\circ; \v{\psi}) = \infty
    \\
    & 
    \implies \Ascr^{\conv,\star}_\sample(\v{\psi}) = -\infty.
\end{aligned}
\end{align}
In other words, existence of a dual feasible point with positive objective implies the existence of a \emph{dual improving ray}, along which the dual objective grows unbounded. In turn, existence of such a ray with a given $\v{\psi}$ certifies that \eqref{eq:socp} is infeasible and $\v{\psi}$ is unacceptable. Assuming that \eqref{eq:socp} is strongly infeasible (see \cite{lourencco2016weak}) for all 
unacceptable $\v{\psi}$, each such $\v{\psi}$ must admit a dual improving ray. Interior point solvers for \eqref{eq:feasibility} typically produce a dual improving ray, when infeasible. See \cite{o2016conic} for the mechanics of producing such a ray. Consider a list of unacceptable configurations $\Uset_\psi := \left\{\v{\psi}^u[1], \ldots, \v{\psi}^u[M']\right\}$. For an unacceptable $\v{\psi}^u[i] \in \Uset_\psi$, let the dual improving ray be given by $\v{\lambda}^\circ[i], \v{\mu}^\circ[i], \v{\mu}_1^\circ[i], \mu_2^\circ[i]$. Then, define
\begin{align}
\begin{aligned}
    \v{\Psi}_{\outS} &= \left\{ \v{\psi} :     \Dcal(\v{\lambda}^\circ[i],\v{\mu}^\circ[i],\v{\mu}_1^\circ[i],\mu_2^\circ[i]; \v{\psi}) \leq 0,
    \right.
    \\
    & \qquad  \qquad \left. i= 1,\ldots,M', \v{\psi} \in [0, \ol{\v{\psi}}] \right\}.
\end{aligned}
\label{eq:def.psi.out}
\end{align}
If $\v{\psi}$ violates at least one among the $M$ affine constraints, then $\v{\psi} \notin \v{\Psi}_\outS$. When \eqref{eq:feasibility} returns a new unacceptable $\v{\psi}$, then $\Uset_\psi \gets \Uset_\psi \cup \v{\psi}$, and the cut defined via its dual improving ray is added to the description of $\v{\Psi}_\outS$ via \eqref{eq:def.psi.out}. Addition of this cut shrinks $\v{\Psi}_\outS$.  

Algorithm \ref{alg:rsasct} crucially relies on the convex nature of the set $\v{\Psi}$--a property that exploits the properties of the \CVaR{} measure. With a probabilistic constraint enforcement as in \cite{geng2020probabilistic}, this set may \emph{not} be convex, rendering the construction of polyhedral inner/outer approximations of $\v{\Psi}$ challenging.

\begin{algorithm}[!t]
    \caption{Incremental Scenario-Based Test $\Ascr^\conv_\sample$.}
    \label{alg:rsasct}
    \begin{algorithmic}[1]
        \State {\textbf{Input: }} $\v{\psi}$, $(\Aset_\psi, \v{\Psi}_\inS), (\Uset_\psi, \v{\Psi}_\outS)$.
        \If{$\v{\psi} \notin \v{\Psi}_\outS$}
            \Return unacceptable.
        \ElsIf{$\v{\psi} \in \v{\Psi}_\inS$}
            \Return acceptable.
        \Else
            \State Solve  \eqref{eq:feasibility}.
            \If{infeasible and dual improving ray $(\v{\lambda}^\circ,\v{\mu}^\circ,\v{\mu}_1^\circ,\mu_2^\circ)$ exists}
                \State $\Uset_\psi \gets \Uset_\psi \cup \v{\psi}$, update $\v{\Psi}_\outS$ via \eqref{eq:def.psi.out}.
                \State \Return unacceptable.
            \Else
                \State $\Aset_\psi \gets \Aset_\psi \cup \v{\psi}$, update             $\v{\Psi}_\inS$ via \eqref{eq:def.psi.in}.
                \State \Return acceptable.
            \EndIf
        \EndIf
    \end{algorithmic}
\end{algorithm}

\begin{figure*}[t]
    \centering
    \includegraphics[width=0.9\textwidth]{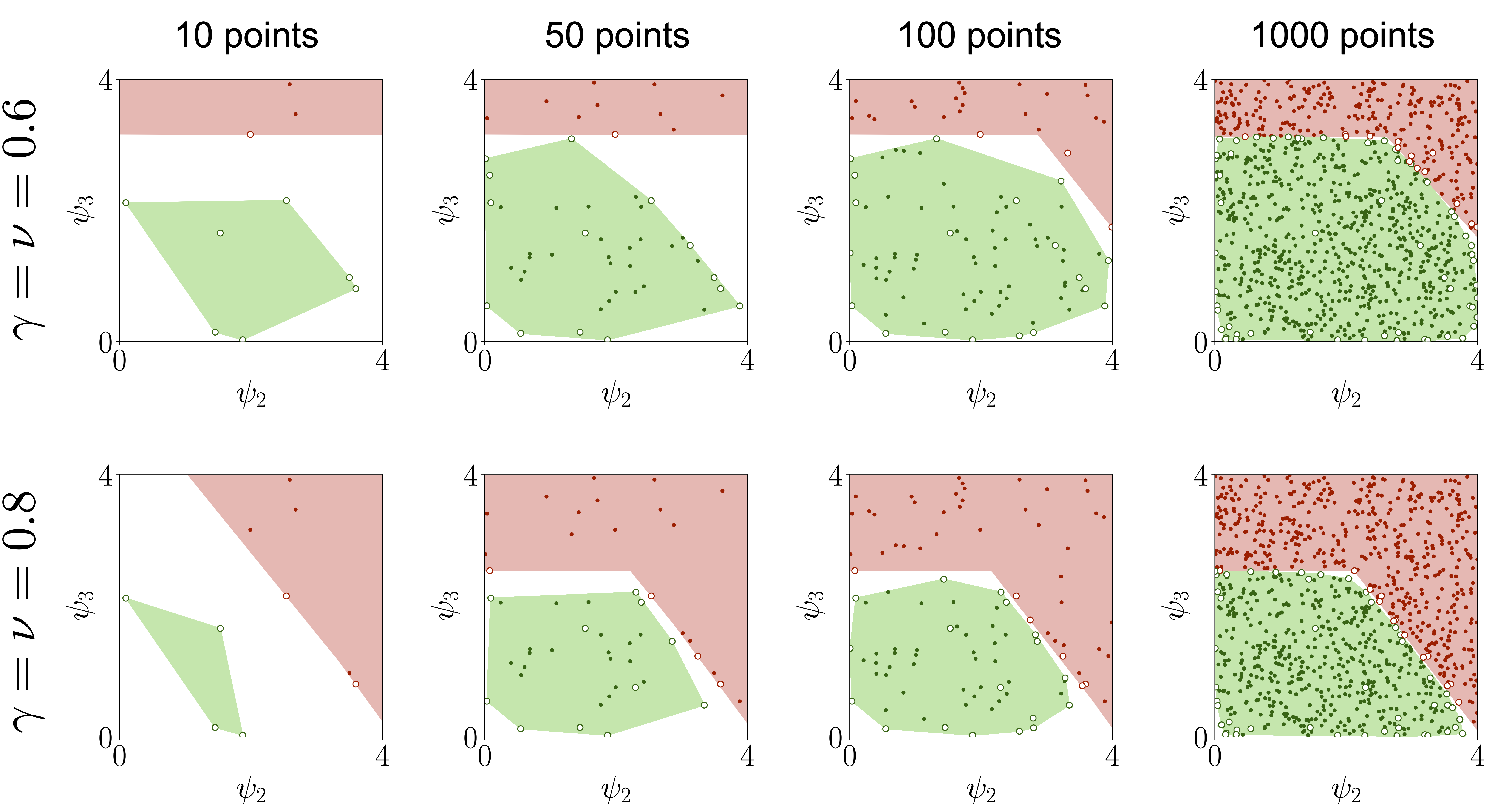}
    \caption{Visualization of $\v{\Psi}_\inS$ (green) and $[0, \ol{\v{\psi}}] \setminus \v{\Psi}_\outS$ (red) regions after testing for acceptability of 10, 50, 100 and 1000 configurations 
    on the 3-bus network example with  
    $\nu = \gamma = 0.6$ on the top and 
    $\nu = \gamma = 0.8$ on the bottom. 
    Hollow points required solving \eqref{eq:feasibility}, while solid points could be certified using $\v{\Psi}_\inS/\v{\Psi}_\outS$.
    } 
    \label{fig:rhsca.case3.cumrun}
\end{figure*}


\subsection{Numerical Experiments with \texorpdfstring{$\Ascr^\conv_\sample(\v{\psi})$}{Apsi}}
\label{sec:feasibility_numerical}
\subsubsection{On the three-bus network} We tested acceptability of uniformly randomly generated installation configurations $(\psi_2, \psi_3) \in [0,4]^2$ p.u. under risk parameters $\nu = \gamma = 0.8$. At the beginning of the testing process, $\v{\Psi}_\inS$ and $\v{\Psi}_\outS$ do not contain relevant information to judge acceptability of candidate configurations, 
and so \eqref{eq:feasibility} gets solved for these configurations. However, as more configurations are tested, $\v{\Psi}_\inS$ grows and $\v{\Psi}_\outS$ shrinks, as Figure \ref{fig:rhsca.case3.cumrun} reveals. As a result, more configurations get quickly certified as acceptable/unacceptable using $\v{\Psi}_\inS$ and $\v{\Psi}_\outS$. As Table \ref{tab:rshca.case3.meanrun} shows, the run-time differences between solving \eqref{eq:feasibility} and testing via $\v{\Psi}_\inS$ and $\v{\Psi}_\outS$ are substantial.

\begin{table}[ht]
    \centering
    \caption{Summary of results from  acceptability tests over the 3-bus distribution network example.}
    \begin{tabular}{l c c}
        \toprule
         Test & Mean Runtime (s) & Frequency  \\ \midrule
         $\v{\psi} \in \v{\Psi}_\inS$ test & $9.678 \times 10^{-4}$ & 451 \\
         $\v{\psi} \notin \v{\Psi}_\outS$ test & $1.195 \times 10^{-6}$ & 489 \\
         Solving \eqref{eq:feasibility}, when $\v{\psi} \in \v{\Psi}_\inS$ & $46.288$ & 47 \\
         Solving \eqref{eq:feasibility}, when $\v{\psi} \notin \v{\Psi}_\outS$ & $133.107$ & 13 \\
         \bottomrule
    \end{tabular}
    \label{tab:rshca.case3.meanrun}
\end{table}
Only a small fraction (6\%) of the 1000 tested configurations, per Table \ref{tab:rshca.case3.meanrun}, needed the solution of \eqref{eq:feasibility} to certify whether they are acceptable. The sets $\v{\Psi}_\inS$ and $\v{\Psi}_\outS$ therefore effectively capture the knowledge of past tests, highlighting the efficacy of our incremental approach to solve $\Ascr^\conv_\sample(\v{\psi})$. 

Figure \ref{fig:rhsca.case3.cumrun} portrays how the sets $\v{\Psi}_\inS$ and $\v{\Psi}_\outS$ change with tests. As one might expect, the set of acceptable configurations with lower risk parameters ($\nu=\gamma=0.6$) in the top half of Figure \ref{fig:rhsca.case3.cumrun} is larger than that with higher risk parameters ($\nu=\gamma=0.8$) in the bottom half of Figure \ref{fig:rhsca.case3.cumrun}.
These figures with increasing number of points were generated with the same set of test configurations, and hence, illustrate how they differently contribute to the construction of the sets $\v{\Psi}_\inS$ and $\v{\Psi}_\outS$ with different risk parameters. 

In practice, installation configurations chosen for certification will likely not be generated uniformly randomly. Rather, these configurations will lie along projected solar adoption paths, starting from an initial acceptable configuration. These capacities being nearby, we expect $\v{\Psi}_\inS$ to be effective in quickly declaring acceptability for most points. Again, as the adoption path leaves $\v{\Psi}$, a few tests near the boundary should construct cutting planes that later points on the trajectory can be quickly ruled out using $\v{\Psi}_\outS$.



\subsubsection{On the 56-bus distribution network}
We ran our tests for acceptable configurations with 15K scenarios over capacity configurations that are uniformly distributed within $[-0.2, 0.2]^5$ p.u. box around $\v{\psi}^\star$ obtained with $\nu=0.9$ and $\gamma = 0.8$. We first ran 32 tests corresponding to the corners of this 5-dimensional hypercube. 
The corner configurations of the hypercube provide valuable information to the construction of $\v{\Psi}_\inS$ and $\v{\Psi}_\outS$. After these 32 corners, we considered an additional $118$ data points sampled uniformly over the aforementioned hypercube. Table \ref{tab:rshca.case56.meanrun} records the results of our numerical experiments. Checking acceptability via $\v{\Psi}_\inS$ or $\v{\Psi}_\outS$ takes less than a second, while running \eqref{eq:feasibility}  takes substantially longer than for the 3-bus network example, owing to the increased network size. Even with a small number (118) of tests, majority of the points are certified to be acceptable or unacceptable using $\v{\Psi}_\inS$ and $\v{\Psi}_\outS$, without requiring explicit solution of \eqref{eq:feasibility}. The run-time differences and the frequency of certification using $\v{\Psi}_\inS$ and $\v{\Psi}_\outS$ illustrate the value of our incremental testing procedure for acceptability of installation configurations for large distribution networks. The gains in run-times can only increase with larger number of scenarios, network sizes, and number of tests conducted.
\begin{table}[!ht]
    \centering
    \caption{Summary of results from  acceptability tests over the 56-bus distribution network example.}
    \begin{tabular}{l c c}
        \toprule
         Test & Mean Runtime (s) & Frequency \\ \midrule
         $\v{\psi} \in \v{\Psi}_\inS$ test & $1.473 \times 10^{-3}$ & 6  \\
         $\v{\psi} \notin \v{\Psi}_\outS$ test & $1.419 \times 10^{-6}$ & 79 \\
         Solving \eqref{eq:feasibility}, when $\v{\psi} \in \v{\Psi}_\inS$ & $978.764$ & 49  \\
         Solving \eqref{eq:feasibility}, when $\v{\psi} \notin \v{\Psi}_\outS$ & $1297.369$ & 16 \\
         \bottomrule
    \end{tabular}
    \label{tab:rshca.case56.meanrun}
\end{table}

\section{Conclusions}
\label{sec:conclusion}

Having already experienced several years of sustained growth, the continued adoption of distributed and utility scale solar generation is in many jurisdictions not only expected, but now mandated by legislation. The primary question facing distribution grid operators and planners has therefore become not if, but how and when solar will be integrated at scale. HCA assesses the limits of a network for safely accommodating solar generation capacity. Given the highly intermittent nature of solar generation, together with changing consumption patterns, moving forward, notions of safety must account for the statistics of both present and forecasted operating conditions. In order to address the scalability issues arising from consideration of large numbers of scenarios, in this work we proposed an HCA formulation based on the use of \CVaR{} as a network constraint risk measure. \CVaR{} provides guarantees as to the frequency and extent of constraint violations, and moreover possesses desirable mathematical properties yielding a tractable approach to HCA. Further leveraging these properties, we provided an expedited method for assessing the acceptability of a given installation configuration, using prior knowledge of installations certified as either acceptable or unacceptable. 

There are a number of interesting research directions we aim to pursue. First, while our optimization approach requires no specialized algorithm design or tuning, the resulting problems are still large and thus, require considerable memory resources. We want to develop decomposition techniques that alleviate such difficulties, without sacrificing quality of the solutions obtained. Second, we aim to explore the potential of scenario reduction, e.g., via clustering, to help relax the computational burden. Finally, we want to model and evaluate the effect of active control of distributed energy resources in CVaR-sensitive HCA with possibly unbalanced three-phase distribution networks.



\bibliographystyle{IEEEtran}
\bibliography{acmart}

\end{document}